# More trouble for unitary quantum theory

J. Finkelstein*

**Abstract**

"Single-world" unitary quantum theories imply that some measurements have results whose probabilities can not be calculated by the Born rule.

The quantum measurement problem: If quantum states evolve unitarily, how can measurements objectively have single outcomes? Well, perhaps quantum states do *not* evolve unitarily (eg the GRW theory [1]). Or perhaps measurements do not have single outcomes (eg many-worlds [2]). Or perhaps quantum states themselves are not objective (eg perspectival states [3] or Quantum Bayesism [4]).

In this note I discuss theories in which it is assumed that quantum states *do* evolve unitarily and that there nevertheless are single, objective results the probabilities of which can be calculated by the Born Rule. I will refer to these as BRUQ theories (for Born-rule-unitary-quantum theories). It has been argued (e.g. in refs. [5] - [9]) that such theories cannot meet the requirements of locality and/or of Lorentz invariance. Without even considering those requirements, I will show that a BRUQ theory suffers from the following trouble: there can be measurement results whose probability cannot be calculated.

Consider a simple story in which Alice and Bob share a laboratory. The entire story will take place in that laboratory, and only a single Lorentz frame will be used; we will not have to consider any constraints from locality or Lorentz invariance. There is in the laboratory a system called S; the initial state of this system is $\frac{1}{\sqrt{2}}$ ($|1\rangle + |2\rangle$). At time t = 1, Alice makes a measurement of S, with result either "1" or "2". Let P(A1) denote the probability that her result is "1". Bob does not see her result; at time t = 3 he makes his own measurement on this same system S; let P(B1) denote the probability that the result of Bob's measurement is "1", and P(A1∧B1) denote the probability that Alice's result and Bob's result are both "1". Let Ψ(t) denote the quantum state vector for the entire laboratory at time t; at t = 0.5, which is before Alice's measurement, Ψ is

$$\Psi(0.5) = \frac{1}{\sqrt{2}} (|1\rangle + |2\rangle) \otimes |Ar\rangle \otimes |Br\rangle \otimes |Er\rangle \qquad (1)$$

where $|Ar\rangle$ is the quantum state vector of Alice before she has made any measurement (but is **r**eady to do so); likewise $|Br\rangle$ is the quantum state vector of Bob before he has made any measurement. The system E includes everything else in the laboratory; $|Er\rangle$ is the quantum state vector of E before any measurement. It follows from eq. (1) and the Born Rule that P(A1) = 0.5.

---
*Lawrence Berkeley National Laboratory
jlfinkelstein@lbl.gov

At t = 1.5, which is after Alice's measurement but before Bob's, Ψ becomes

$$\Psi(1.5) = \frac{1}{\sqrt{2}} \ [(|1\rangle \otimes |A1\rangle \otimes |E1\rangle )+(|2\rangle \otimes |A2\rangle \otimes |E2\rangle)] \otimes |Br\rangle \qquad (2)$$

where $|A1\rangle$ ($|A2\rangle$) are state vectors of Alice where she has seen results "1" (resp. "2"); with analogous meanings for $|E1\rangle$ ($|E2\rangle$ ). It follows from eq. 2 and the Born rule that P(B1) = 0.5. Of course, the values of P(A1) and P(B1) do not determine the value of P(A1∧B1) (unless either P(A1) or P(B1) were zero, which is not the case here). That value can be determined from eq. (2) by considering a measurement on the combined system of S and Alice of a observable whose associated operator has $|1\rangle \otimes |A1\rangle$ as an eigenvector. The Born rule then implies that P(A1∧B1)= 0.5.

So we have P(A1) = 0.5, P(B1) = 0.5, and P(A1∧B1) = 0.5. So far, there is no trouble. But consider a second version of our story: it's the same as the first version, except that at time t = 2 the quantum state of the laboratory is reset to its original value; that is, the laboratory undergoes a unitary evolution which is the inverse of the evolution that occurred during Alice's measurement. Of course that would be impossible to implement in practice (a similar story in ref [9] invokes a "superobserver' with the power to reset the quantum state of a macroscopic system) but it is allowed in principle by the rules of unitary quantum theory.

I will use the letter Φ to denote quantum state vectors in this second version, to avoid confusion with state vectors in the first version as given above. Before Alice's measurement the state vector for the laboratory is the same as in the first version (eq. 1):

$$\Phi(0.5) = \frac{1}{\sqrt{2}} \ (|1\rangle+|2\rangle)\otimes|Ar\rangle\otimes|Br\rangle\otimes|Er\rangle \qquad (3)$$

From this and the Born rule, we see that P(A1) = 0.5.

At t = 1.5, which is after Alice's measurement but before the reset, the state vector is the same as in the first version (eq. 2):

$$\Phi(1.5) = \frac{1}{\sqrt{2}} \ [(|1\rangle \otimes|A1\rangle\otimes|E1\rangle)+(|2\rangle\otimes|A2\rangle\otimes|E2\rangle)]\otimes|Br\rangle \qquad (4)$$

However, we learn nothing about Bob's result from eq. (4). The Born rule, applied to Φ(1.5), would give the probability of a measurement which is made when the state vector is Φ(1.5) , but that is not what Bob does. We can calculate P(B1) from the state vector at t = 2.5, which, due to the reset, is the same as it was at t=0.5:

$$\Phi(2.5) = \frac{1}{\sqrt{2}} \ (|1\rangle+|2\rangle)\otimes|Ar\rangle\otimes|Br\rangle\otimes|Er\rangle \qquad (5)$$

and we see that P(B1) = 0.5 (same as in the first version).

The trouble is that there is now no way to use the Born rule to calculate P(A1∧B1). Eq. 4 tells us nothing about Bob's result, and eq. 5 tells us nothing about Alice's result. One might want to conjecture that, since there is no obvious reason for Alice's result and Bob's result to be correlated, they should be statistically independent; that would imply that P(A1∧B1) = 0.25. Unfortunately, that conjecture is not supported by any Born rule calculation, and in fact is, as we shall see below, contradicted by what we would expect from Bohmian mechanics.

In Bohmian mechanics (for a summary of Bohmian mechanics, see ref. [10]), the quantum state vector is supplemented by an actual position for the particle, which I will call the "Bohm position"; the dynamics of the Bohm position is such that the probability of the Bohm position being in a certain place is equal to the probability that a measurement of position would find it there. For our story, let the system S be a particle which could be in one of two boxes; $|1\rangle$ is the state vector of the particle if the particle is in box #1, and $|2\rangle$ the state vector if the particle is in box #2. With the usual choice of a measurement Hamiltonian, a wave packet which starts in box #1 will never overlap in space with a wave packet which starts in box #2; that implies that the Bohm position will not travel from one box to the other. With probability 0.5 the Bohm position will start in box #1, in which case it would stay there for the entire duration of the story, and so both Alice and Bob would see result "1". Thus P(A1∧B1) = 0.5 (as in the first version). This is not in conflict with the assertion that a BRUQ theory does not permit a calculation of P(A1∧B1). This Bohmian calculation relies on the postulated dynamics of the Bohm position, not on the Born rule; therefore Bohmian mechanics does not meet the definition I have made of a BRUQ theory. The "trouble" in the title of this note does not occur in Bohmian mechanics; it would occur in any unitary theory in which probabilities are just those that follow from the Born rule.

Final observation: In many interpretations of quantum theory, probabilities can only be assigned to observed results. In the second version of the story I have just told, after the quantum state is reset Alice's result is not observable; all records of it, including any in Alice's memory, have been erased. So perhaps P(A1∧B1) should not be considered to be a meaningful quantity. But to the extent that it *is* meaningful, it is an example of a probability to which a BRUQ theory does not assign a value.

Acknowledgment: I acknowledge the hospitality of the Lawrence Berkeley National Laboratory, where this work was done.